\documentclass[twoside]{article}
\usepackage{fleqn,espcrc2}

\usepackage{amssymb}
\usepackage{epsfig}
\usepackage{eucal}
\usepackage{amsmath}

\renewcommand{\cal}{\CMcal}                  % STILE CALLIGRAFICO
\newcommand{\pa}{\mbox{\scriptsize\sf a}}    % PARTICELLA A
\newcommand{\pb}{\mbox{\scriptsize\sf b}}    % PARTICELLA B
\newcommand{\pq}{\mbox{\scriptsize\sf q}}    % PARTICELLA QUARK
\newcommand{\pg}{\mbox{\scriptsize\sf g}}    % PARTICELLA GLUONE
\newcommand{\as}{\alpha_s}                   % COSTANTE FORTE 
\newcommand{\ab}{\overline{\alpha}_s}
\newcommand{\kk}{{\boldsymbol k}}            % VETTORI TRASVERSI
\newcommand{\ku}{{\boldsymbol k}_1}
\newcommand{\kd}{{\boldsymbol k}_2}
\newcommand{\qq}{{\boldsymbol q}}
\newcommand{\dif}{{\rm d}}                   % DIFFERENZIALI
\newcommand{\du}{\dif[\ku]}\newcommand{\dd}{\dif[\kd]}

\newcommand{\G}{{\cal G}}                    % FUNZIONE DI GREEN
\newcommand{\K}{{\cal K}}                    % QUANTITA' UNIVERSALE
\newcommand{\e}{\varepsilon}                 % DIMENSIONI EXTRA
\renewcommand{\o}{\omega}
\newcommand{\g}{\gamma}                   % \gamma
\newcommand{\eq}[1]{eq.(\ref{#1})}         
         
\newcommand{\beq}{\begin{equation}}
\newcommand{\eeq}{\end{equation}}
\newcommand{\bea}{\begin{eqnarray}}
\newcommand{\eea}{\end{eqnarray}}
\newcommand{\non}{\nonumber}

\title{QCD factorization with heavy quarks}

\author{Germ\'an Rodrigo~\address{
Institut f\"ur Teoretische Teilchenphysik,
Universit\"at Karlsruhe, D-76128 Karlsruhe, Germany.}
\thanks{Supported by INFN (Italy) and BMBF Project 05HT9VKB0 (Germany).
Email: rodrigo@particle.uni-karlsruhe.de} and
        Marcello Ciafaloni~\address{ 
Dipartimento di Fisica, Universit\`a di Firenze and INFN 
Sezione di Firenze, Largo E. Fermi 2, I-50125  Firenze, Italy.}
\thanks{Supported by E.U. QCDNET contract
FMRX-CT98-0194 and MURST (Italy).
Email: ciafaloni@fi.infn.it}}
       
\begin{document}

\begin{abstract}
We further analyze the definition and the calculation of
the heavy quark impact factor at next-to-leading (NL)
$\log s$ level, and we provide its analytical expression
in a previously proposed $\kk$-factorization scheme. 
Our results indicate that $\kk$-factorization holds at NL level
with a properly chosen energy scale, and with the same gluonic
Green's function previously found in the massless probe case.
\end{abstract}

%\preprint{DFF367/10/2000 \\ TTP00-23}

\maketitle

\section{INTRODUCTION}

Recent improvements~\cite{s98} of the next-to-leading 
$\log x$ (NL$x$) results~\cite{bfkl98}
in the BFKL framework, have stabilized the small-x 
behaviour in QCD, so that a phenomenological analysis 
of deep inelastic processes (DIS) seems now possible.

However, both the gluon density (satisfying the improved equation)
and the impact factors are needed in order to use 
$\kk$-factorization~\cite{c98} to compute DIS or double DIS processes. 
So far, NL$x$ impact factors have been found for the unphysical 
case of massless initial quarks and gluons only~\cite{c98,impact99}. 
Partial features for massive quarks~\cite{pro,fmassive99} 
and for colourless sources~\cite{fcolorless99} are known too.

In this talk we present the complete results for the case of initial 
massive quarks derived in Ref.~\cite{heavy}. These results allowed us 
to check the validity of the $\kk$-factorization scheme introduced in 
Ref.~\cite{impact99}, or, in other words, to derive probe independent 
gluon Green's function with an explicit massive quark impact factor
which satisfies the expected collinear properties. 
Furthermore, we developed as a byproduct some analytical 
techniques which are needed to deal with two-scale problems,
which are hopefully useful to cope with the physical cases also.

\section{k-FACTORIZATION IN DIJET PRODUCTION}

\label{sec:kfactor}

\begin{figure}
\begin{center}
\epsfig{file=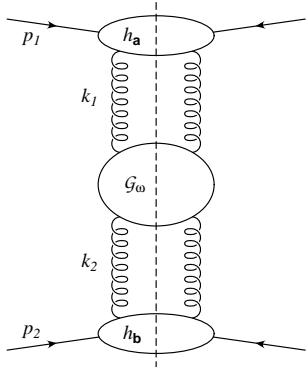,width=4cm}
\end{center}
\caption{Diagrammatic representation of $\kk$-factorization.}
\label{fig:fact}
\end{figure}

Following~\cite{c98}, the colour averaged differential cross section
for the high-energy scattering of two partons {\sf a} and {\sf b}
is factorized in a gauge-invariant way into a Green's function $\G_{\o}$
and impact factors $h_{\pa}$ and $h_{\pb}$ (Fig.~\ref{fig:fact})
\begin{align}
 \frac{\dif\sigma_{\pa\pb}}{\du\,\dd} =
  \int & \frac{\dif\o}{2\pi i\o}\, 
  \left(\frac{s}{s_0(\ku,\kd)}\right)^{\o} \label{fatt} \\ & \times 
  h_{\pa}(\ku) \G_{\o}(\ku,\kd) h_{\pb}(\kd)~. \non
\end{align}
The transverse momenta  $\ku$ and $\kd$, defined with respect
to the incoming momenta $p_1$ and $p_2$, play the role of
hard scales of the process. 

At the next-to-leading $\log x$ (NL$x$) accuracy 
the Green's function $\G_{\o}$ has the following general form 
\begin{align}
 \G_{\o} = (1+\ab H_L) & \left[1-\frac{\ab}{\o}(K_0+K_{NL})\right]^{-1}\!
 \non \\ & \times (1+\ab H_R)~,
 \label{scomp}
\end{align}
where $K_0$ and $K_{NL}$ are the leading $\log x$ (L$x$) and the
NL$x$ BFKL kernels~\cite{bfkl98} respectively,
$H_R (H_L)$ are operator factors introduced in~\cite{impact99}
so as to provide partonic impact factors free of double $\log$ 
collinear divergences and $\ab = \as N_c/\pi$ is the dimensionless 
strong coupling constant.

As explained in~\cite{impact99}, the identification of the second
order impact factors, $h_{\pa}^{(1)}$ and $h_{\pb}^{(1)}$,
is affected by a double factorization scheme ambiguity, due to both 
the choice of the scale $s_0$ and of the kernels $H_R (H_L)$.

\section{FACTORIZATION SCHEME AND CALCULATIONAL PROCEDURE}

\begin{figure}
\begin{center}
\epsfig{file=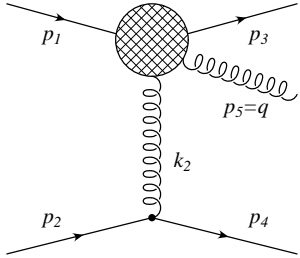,width=4cm}
\end{center}
\caption{Real gluon emission in the fragmentation region of quark ${\sf a}$.}
\label{fig:frag}
\end{figure}

Let's consider first the high-energy scattering of
two partons ${\sf a}$ and ${\sf b}$ where
${\sf a}={\sf q}$ is a heavy quark of mass $m$ with real
emission of an extra gluon ${\sf g}$ that we assume in the
heavy quark fragmentation region (Fig.~\ref{fig:frag}).
In terms of invariants $s_2=(q+p_4)^2 \gg s_1=(p_3+q)^2$.
The Born differential cross section in this high energy region 
was calculated in~\cite{heavy}. Though complicated at first sight
it reduces, as expected to the known~\cite{impact99}
result for $m \rightarrow 0$, and matches the 
L$x$ differential cross section 
\begin{align}
\frac{\dif\sigma^{(L)}_{\pq\pg\pb}}{\dif z_1\,\du\,\dd} & =
h_{\pq}^{(0)}(\ku) \, h_{\pb}^{(0)}(\kd)  \non \\ & \times
\frac{\ab}{\qq^2\Gamma(1-\e)\mu^{2\e}} \, \frac{1}{z_1}~,
\label{Lxreal}
\end{align}
in the limit $z_1 \rightarrow 0$, being $z_1$ the momentum fraction
of $k_1=p_1-p_3$ with respect to the incoming momentum $p_1$,
$h_{}^{(0)}({\kk_i})$ the leading order impact factor and $\qq=\ku+\kd$.

However, as pointed out in~\cite{heavy}, for~\eq{Lxreal} to
be a good approximation to the total result, we should require 
($q=|\qq|,k_i=|\kk_i|$)
\begin{equation}
z_1 \ll q/k_1~, \quad  q/m~, \quad k_1/m~.
\label{constrains}
\end{equation}
The first two cutoffs can be summarized by $z_1 < q/Max(k_1,m)$, which
is a coherence condition for the case of heavy quarks, saying that 
the rapidity of the gluon cannot exceed that of the final quark. 

By integrating the leading expression~(\ref{Lxreal}) with the 
constraints~(\ref{constrains}) in the fragmentation region 
$z_1>q/\sqrt{s}$, an estimate of the leading contribution contained in
the complete result, which should be subtracted out in order to yield 
the impact factor in the massive quark case, was found~\cite{heavy}.
Then, by considering both real and virtual contributions to the 
fragmentation function $F_{\pq}(z_1,\ku,\kd)$, we introduced the 
following definition of the impact factor $h_{\pq}^{(1)}(\kk)$:
\begin{align}
& \int_{q/\sqrt{s}}^1 \dif z_1 \int \du 
F_{\pq}(z_1,\ku,\kd)  = \non  \\
&= h_{\pq}^{(1)}(\kd) + 
\int \du \, \ab \, h_{\pq}^{(0)}(\ku) \, K_0(\ku,\kd) \, \non \\
& \qquad \times \left( \log \frac{\sqrt{s}}{Max(k_1,m)} 
- \log \frac{q}{k_1} \Theta_{q k_1} \right)~. 
\label{h1massive}
\end{align}

\label{sec:scale}

Compared to the subtraction (or factorization) scheme adopted 
in~\cite{impact99} for $m=0$, the expression~(\ref{h1massive}) differs 
by the replacement $k_1 \rightarrow Max(k_1,m)$, which leads,
by adding the symmetrical fragmentation region, to the choice 
for the factorized scale in~\eq{fatt}
\begin{equation}
s_0=Max(k_1,m_1)Max(k_2,m_2)~,
\label{scale}
\end{equation}
$m_1$ being the mass of quark ${\sf a}$ 
and $m_2$ the mass of quark ${\sf b}$,
and in particular contains the 
subtraction term $\log q/k_1 \Theta_{q k_1}$ which
provides the expression ($H_R=H_L^{\dagger}=H$)
\begin{equation}
 H(\ku,\kd)=-\frac{1}{\qq^2\Gamma(1-\e)\mu^{2\e}}\,\log\frac{q}{k_1}\;
 \Theta_{q k_1}~, 
 \label{acca}
\end{equation}
for the $H$ kernel in the $\kk$-factorization formula. 

In order to simplify the subsequent calculations, 
the known result~\cite{impact99} for $m=0$ was used 
and only the difference for a non vanishing mass 
\begin{align}
\Delta F_{\pq}& (z_1,\ku,\kd) = \non \\ & =
F_{\pq}(z_1,\ku,\kd) - F^{m=0}_{\pq}(z_1,\ku,\kd)~,
\label{splitF}
\end{align}
was explicitly computed. Then, we found the following relationship 
between the massless quark and the heavy quark impact factors
\begin{align}
h_{\pq}^{(1)}&(\kd) = h_{\pq,m=0}^{(1)}(\kd) + \label{h1mass} \\ 
& + \int_0^1 \dif z_1 \int \du \Delta F_{\pq}(z_1,\ku,\kd) \non \\
& + \int \du \, \ab \, h_{\pq}^{(0)}(\ku) \, K_0(\ku,\kd) \,
  \log \frac{m}{k_1} \, \Theta_{m \, k_1}~. \non 
\end{align}
Notice that the integration limits in $z_1$ have been extended
down to $z_1=0$. Since $\Delta F_{\pq}$ is regular at $z_1=0$ this 
change introduces only a negligible error of order $1/s$.

\section{MELLIN TRANSFORM AND ITS INVERSE}

In order to perform the calculation outlined in~\eq{h1mass},
we proceeded in two steps. First, the $\ku$ integration was 
performed analytically by reducing the $\ku$-integrals to two 
denominators. Then, the virtual contribution~\cite{ffqk96}
was considered and organized in terms of momentum fraction 
integrals only. Finally, summing up real and virtual contributions 
to the fragmentation vertex we obtained an expression for the 
difference $\Delta F_{\pq}(\kd)$, arising from the second term in
the r.h.s. of~\eq{h1mass}. To perform the last integrations we 
calculated its Mellin transform 
\begin{align*}
\Delta \tilde{F}_{\pq}(\gamma) &=  
\Gamma(1+\e) \, (m^2)^{-\e} \non \\ & \times
\int \dd  \left( \frac{\kd^2}{m^2}\right)^{\g-1}
\Delta F_{\pq}(\kd)~,  
\end{align*}
which allowed us to disentangle the $(m/k)$-dependence, yielding
\begin{align}
\Delta & \tilde{F}_{\pq}(\gamma)  = A_{\e} \, (m^2)^{\e} \non \\
& \times \frac{\Gamma(\g+\e)\Gamma(1-\g-2\e)\Gamma^2(1-\g-\e)}
 {8\Gamma(2-2\g-2\e)} \non \\
& \times 
\bigg[ \frac{1+\e}{\g+2\e} + \frac{2}{1-2\g-4\e} \non \\ & \qquad \times
\left( \frac{1}{1-\g-2\e}- \frac{1}{3-2\g-2\e} \right) \bigg]~,
\label{mellin}
\end{align}
where $A_{\e}$ is a constant that contains the dependence 
on the strong coupling constant and some colour factors. 

It is straightforward, though not trivial, to show 
that~\eq{mellin} converges only in the small band 
$1-2\e < Re \g <1-\e$. The inverse Mellin transform 
was thus defined as 
\begin{align*}
\Delta F_{\pq}(\kd) =
\frac{1}{m^2} & \int_{1-2\e < Re \g <1-\e} \frac{\dif \g}{2\pi i}  
\\ & \times
\left(\frac{\kd^2}{m^2}\right)^{-\g-\e}
\Delta \tilde{F}_{\pq}(\gamma)~.  
\end{align*}
Then, displacing the integration contour around the positive 
or the negative real semiaxis, i.e. enclosing all the poles
placed either at $\g \ge 1-\e$ or $\g \le 1-2\e$, we calculated 
the different corrections of order $\cal{O}(m/k_2)^n$ 
or $\cal{O}(k_2/m)^n$ to the impact factor in the limits 
$\kd^2>m^2$ and $\kd^2<m^2$ respectively.

\section{IMPACT FACTOR}

\label{sec:impact}

Our final result for the heavy quark impact factor 
at the next-to-leading level reads
\begin{align}
& h_{\pq}(\kd) =  h_{\pq}^{(1)}(\kd) \big|_{sing} +
 h_{\pq}(\kd) \big|_{finite}~,
\end{align}
where the singular piece is defined as
\begin{align}
 h_{\pq}^{(1)}&(\kd) \big|_{sing} 
= \delta h_{1}^{(1)}(\kd) \label{h1singular} \\
&+ h_{\pq}^{(0)}(\kd) \, \frac{\as N_c}{2\pi}
\left( - \frac{3}{2} \log \frac{\kd^2}{m^2} \right) \Theta_{k_2 \, m}~, \non 
\end{align}
and 
\begin{align}
 h_{\pq}&(\kd) \big|_{finite}
= h_{\pq}^{(0)}(\as(\kd))  \bigg\{ 1+\frac{\as N_c}{2\pi} 
\non \\ \times & \bigg[
\K - \frac{\pi^2}{6} - \bigg(\frac{3}{2} + 
\sum_{Re\g >1} Res [\tilde{h}(\g)] \bigg) \Theta_{k_2 \, m}
\non \\
& + \bigg( 2 +  
\sum_{Re\g <1} Res [\tilde{h}(\g)] \bigg) \Theta_{m \, k_2}
\bigg] \bigg\}~,
\label{final}
\end{align}
is the finite contribution, with
\begin{equation}
\K = \frac{67}{18}-\frac{\pi^2}{6}-\frac{5n_f}{9N_c}~.
\end{equation}
As for the massless case, the singularities proportional to
$(11/6-n_f/3N_c)$, the beta function, were absorbed by the running
strong coupling constant $\as(\kd)$.
The function $\tilde{h}(\g)$ provides the corrections~\cite{heavy} 
of order $\cal{O}(m/k_2)$ and $\cal{O}(k_2/m)$ to the impact factor 
for $\kd^2>m^2$ and $\kd^2<m^2$ respectively.

Notice that all double $\log$ contributions
of type $1/\e^2$ and $1/\e \log(\kd^2/m^2)$ appearing in
the intermediate steeps of the calculation canceled out which
means that indeed our subtraction of the leading kernel was
effective, thus lending credit to the scale~(\ref{scale})
and to the $H$-kernel~(\ref{acca}).
The remaining singularities of the impact factor are single
logarithmic ones~$\sim~1/\e$. In fact, the impact factor is actually
{\it finite}, with the expected $\log(\kd^2/m^2)$ dependence predicted
by the DGLAP equations, the divergent piece
$\delta h_{1}^{(1)}(\kd)$ in \eq{h1singular}, see~\cite{heavy},
can be interpreted as a finite mass scale change, i.e.
the scale leading to a finite massive quark impact factor differs 
from~\eq{scale} by a finite renormalization of the quark mass, 
which is a normal ambiguity in this type of problems. 

\section{CONCLUSIONS}

\label{sec:conclusion}

Starting from the explicit squared matrix element for gluon
emission we motivated the subtraction of the
leading term, and we performed the $\ku$
and $z_1$ integrals needed to provide an explicit result for the 
heavy quark impact factor.

Even if the cross section being investigated is unphysical,
the relevance of our results stems from the consistency of the
following features: (i) the validity of the $\kk$-factorization
formula~(\ref{fatt}) with scale $s_0=Max(k_1,m_1)Max(k_2,m_2)$;
(ii) the explicit expression of the impact factor with
factorizable single logarithmic collinear divergences, and
(iii) the probe-independence of the subleading $H$-kernels 
of the CC scheme~\cite{impact99}, defined in \eq{acca}.

Of course, the real problem is to provide an
explicit expression for the DIS impact factors. But -- if the
lesson learned form the L and NL kernels is still valid --
the impact factor's magnitude is not expected to be much 
different from their approximate collinear evaluation.

%%%%%%%%%%%  BIBLIOGRAFIA  %%%%%%%%%%%%

\end{document}